%% file: main.tex
\documentclass[copyright,creativecommons]{eptcs}
\input{header}

\title{An Overview of the Decentralized Reconfiguration Language Concerto-D through its Maude Formalization}
\author{Farid Arfi \quad\qquad Hélène Coullon
\institute{IMT Atlantique, Inria, LS2N, UMR 6004, F-44000 Nantes, France}
\email{\{farid.arfi, helene.coullon\}@imt-atlantique.fr}
\and Frédéric Loulergue
\institute{Univ. Orléans, INSA CVL, LIFO EA 4022, France}
\email{frederic.loulergue@univ-orleans.fr}
\and Jolan Philippe
\institute{IMT Atlantique, Inria, LS2N, UMR 6004, F-44000 Nantes, France}
\email{jolan.philippe@imt-atlantique.fr}
\and Simon Robillard
\institute{LIRMM, CNRS, Université de Montpellier, France}
\email{simon.robillard@umontpellier.fr}
}

\def\titlerunning{The Decentralized Reconfiguration Language Concerto-D}
\def\authorrunning{Arfi et. al.} 

\begin{document}
\maketitle

\begin{abstract}
We propose an overview of the decentralized reconfiguration language \concertod through its Maude formalization. \concertod extends the already published \concerto language. \concertod improves on two different parameters compared with related work: the decentralized coordination of numerous local reconfiguration plans which avoid a single point of failure when considering unstable networks such as edge computing, or cyber-physical systems (CPS) for instance; 
and a mechanized formal semantics of the language with Maude which offers guarantees on the executability of the semantics. Throughout the paper, the \concertod language and its semantics are exemplified with a reconfiguration extracted from a real case study on a CPS. We rely on the Maude formal specification language, which is based on rewriting logic, and consequently perfectly suited for describing a concurrent model. 
\end{abstract}


\section{Introduction}
\label{sec:intro}
\input{Introduction}
\section{Control components and reconfiguration language}
\label{sec:components}
\input{sections/components}

\section{Elements of operational semantics}
\label{sec:semantics}
\input{sections/full_example}
\subsection{Execution and synchronization of reconfiguration actions}
\label{sec:firing}
\input{sections/semantics}
%
\subsection{Communications}
\label{sec:communications}
\input{sections/communications}

\section{Related work}
\label{sec:RelatedWork}
\input{RelatedWork}
\section{Conclusion and Future Work}
\label{sec:Conclusion}
\input{Conclusion}




\bibliographystyle{eptcs}
\bibliography{main}
\end{document}

%% file: header.tex
\usepackage{amsmath,amssymb,amsfonts} 
\usepackage[normalem]{ulem}
\usepackage[inkscapeformat=pdf]{svg}
\usepackage{lscape} 
\usepackage{graphicx}
\usepackage{hyperref}
\usepackage{multirow}
\usepackage{subfig}
\usepackage{xcolor,pifont}
\usepackage[textwidth=17mm]{todonotes}
\usepackage{listings}
\usepackage{xspace}
\usepackage[numbers]{natbib}
\usepackage{doi}
\hypersetup{hidelinks}
\usepackage[utf8]{inputenc}
\usepackage[T1]{fontenc}
\usepackage[inline]{enumitem}
\usepackage{svg}

\usepackage{multicol}

\input{_maude}
\input{_concerto}

\newcommand{\tool}[1]{\textsc{#1}\xspace}

\newcommand{\concerto}{\tool{Concerto}}
\newcommand{\concertod}{\tool{Concerto-D}}
\newcommand{\puppet}{\tool{Puppet}}

\newcommand{\mupuppet}{$\mu$\tool{Puppet}}
\newcommand{\smartfrog}{\tool{SmartFrog}}

\usepackage{xspace}
\newcommand{\ie}[0]{i.e.,\xspace}

\newcommand{\eg}[0]{e.g.,\xspace}
\newcommand{\etal}[0]{et al.\xspace}
\newcommand{\wrt}[0]{w.r.t.\xspace}

\lstdefinelanguage{Python}{
morekeywords=[5]{gateway,system},
keywordstyle=[5]{\ttfamily\textbf},
}

\lstMakeShortInline[language=Python]"

\definecolor{keywords}{RGB}{255,0,90}
\definecolor{comments}{RGB}{0,0,113}
\definecolor{red}{RGB}{160,0,0}
\definecolor{green}{RGB}{0,150,0}
\definecolor{backcolour}{rgb}{0.95,0.95,0.95}
\DeclareCaptionFont{white}{\color{white}}
\DeclareCaptionFont{gray}{\color{gray}}
\DeclareCaptionFormat{listing}{\colorbox{backcolour}{\parbox{\linewidth}{#1#2#3}}}
\usepackage{iftex}

\ifpdf
  \usepackage{underscore}         
\else
  \usepackage{breakurl}           
\fi
\usepackage[english]{babel}

\usepackage{tikz}
\usetikzlibrary{positioning, arrows.meta, decorations.pathreplacing, fit}
\pgfarrowsdeclare{lollipop}{lollipop}{
    \pgfsetlinewidth{0.7pt}
    \pgfarrowsleftextend{0pt}
    \pgfarrowsrightextend{2pt}
}{
    \pgfpathcircle{\pgfpoint{0pt}{0pt}}{2pt}
    \pgfusepathqfillstroke
}

\pgfarrowsdeclare{socket}{socket}{
    \pgfarrowsleftextend{2pt}
    \pgfarrowsrightextend{2pt}
}{
    \pgfsetlinewidth{0.7pt}
    \pgfpathmoveto{\pgfqpoint{3pt}{-3pt}}
    \pgfpatharc{270}{90}{3pt}
    \pgfusepathqstroke
}

\tikzset{
  lollipop/.style={
    -{lollipop}
  },
  socket/.style={
    -{socket}
  }
}

\newcommand{\provideusetwolabels}[5][1cm]{%
    \path (#2) -- (#3) coordinate[pos=0.5] (mid#2#3);
    \path (#2) -- (mid#2#3) coordinate[pos=0.5] (q1#2#3);
    \path (mid#2#3) -- (#3) coordinate[pos=0.5] (q3#2#3);
    \draw[-{lollipop}] (#2) --  node[midway, above, xshift=0.08cm] {\tiny #4} (q1#2#3);
    \draw[-{lollipop}] (mid#2#3) -- (q3#2#3);
    \draw[-{socket}] (#3) -- node[midway, below, xshift=-0.4cm] { \tiny #5} (q3#2#3);
    \draw[-{socket}] (q3#2#3) -- (q1#2#3);
}

\newcommand{\provideusesinglelabel}[4][1cm]{%
    \path (#2) -- (#3) coordinate[pos=0.5] (mid#2#3);
    \path (#2) -- (mid#2#3) coordinate[pos=0.5] (q1#2#3);
    \path (mid#2#3) -- (#3) coordinate[pos=0.5] (q3#2#3);
    \draw[-{lollipop}] (#2) -- (q1#2#3);
    \draw[-{lollipop}] (mid#2#3) -- (q3#2#3);
    \draw[-{socket}] (#3) -- (q3#2#3);
    \draw[-{socket}] (q3#2#3)  -- node[midway, above] {\scriptsize #4} (q1#2#3);
}

%% file: _maude.tex
\definecolor{MaudeDelimiterColor}{HTML}{B65E47}
\definecolor{MaudeNumberColor}{HTML}{FF0000}
\definecolor{MaudeCommentColor}{HTML}{008000}
\definecolor{MaudeKeyColor}{HTML}{002BFF}
\definecolor{MaudeSymbolColor}{HTML}{0030FF}

\lstdefinelanguage{maude}
{
	numbers=left,
    numberstyle=\tiny\color{gray}\sffamily,
	breaklines=true,
	extendedchars=true,
	tabsize=2,
	columns=[l]flexible,
	showtabs=false,
	showstringspaces=false,
	showspaces=false,
	showstringspaces=false,
    basicstyle=\ttfamily\small,
	keywordstyle=\bfseries\color{MaudeKeyColor},
	ndkeywordstyle=\color{MaudeKeyColor},
	stringstyle=\color{MaudeDelimiterColor},
	commentstyle=\itshape\rmfamily\color{MaudeCommentColor},
    keywordstyle=[2]{\ttfamily\bfseries\color{red}},
    morekeywords=[2]{<,>,=>},
	ndkeywords={Int, Bool, String},
	keywords={in, load, pr, protecting, sort, sorts, inc, op, ops, var, vars, eq, cq, ceq, crl, ctor, endfm, fmod, endm, is, mod, endm, load, =, ==, =/=, true, false, nil, empty, if, then, else, class, load, such, that, subclass, Prop, omod, sort, subsort, not, endom, fi, and, or, endtm, using, assoc, comm, view, endv, from, to, subsorts, rl},
	morecomment={[l]{***}},
	morecomment={[l]{---}},
    literate=
    {=>}{{\color{MaudeSymbolColor}{\;{=}{>}\;}}}2
    {->}{{\color{MaudeSymbolColor}{\;->\;}}}2
    {<}{{\color{MaudeSymbolColor}< }}1
    {>}{{\color{MaudeSymbolColor}> }}1
    {|->}{{\color{MaudeSymbolColor}{$\mapsto$} }}3
}

%% file: _concerto.tex
\definecolor{ConcertoStringColor}{HTML}{B65E47}
\definecolor{ConcertoNumberColor}{HTML}{FF0000}
\definecolor{ConcertoCommentColor}{HTML}{008000}
\definecolor{ConcertoCmdColor}{HTML}{0010AA}
\definecolor{ConcertoCFlowColor}{HTML}{00AA00}

\lstdefinelanguage{concertod}
{
	breaklines=false,
	extendedchars=true,
    mathescape=true,
	tabsize=2,
	columns=[l]flexible,
	showtabs=false,
	showstringspaces=false,
	showspaces=false,
	showstringspaces=false,
    basicstyle=\ttfamily,
	keywordstyle=\bfseries\color{ConcertoCmdColor},
    keywordstyle=[2]\bfseries\color{ConcertoCFlowColor},
	stringstyle=\color{ConcertoStringColor},
	commentstyle=\itshape\rmfamily\color{ConcertoCommentColor},
    keywords={pushB,wait},
    keywords=[2]{for,in,range},
}

%% file: Introduction.tex
Running and maintaining large-scale distributed software is now a commonplace activity, but managing the inherent complexity of this task requires dedicated tools, models, and languages. The complexity is particularly apparent when distributed software needs to be reconfigured during execution, either to satisfy changing requirements or to carry out maintenance operations. 

The DevOps community (and associated tools) as well as component-base software engineering (CBSE) are the main domains focussing on the deployment and reconfiguration of distributed software systems. A reconfiguration consists of a set of actions to execute on the different pieces of software, distributed across the network, to lead the system in the new desired configuration (\ie state). In these domains, actions are almost always orchestrated by a central coordinator~\cite{Sokolowskietal, CHL2023:CSUR}, \ie an entity that keeps track of the actions to apply and their dependencies, as well as the global state of the distributed system.

However, a centralized model is necessarily limited in terms of resilience, as it creates a single point of failure. 
For instance, in the context of constrained (\eg energy, communications) cyber-physical systems~\cite{omond:hal-04372320,omond:hal-04372340}, or edge computing where network disconnections are commonplace, as well as in the context of large-scale projects where cross-DevOps teams~\cite{6727828, PhilippeOCPR2024:SANER, Sokolowskietal, wildetal} have to collaborate, decentralized reconfiguration models are preferred.

With this work, we extend the semantics of the reconfiguration language \concerto~\cite{CHARDET2021102582} and turn it into a decentralized model called \concertod, by extending the semantics to describe the specifics of communication and synchronization between components. In \concertod multiple coordinators collaborate to achieve their respective local reconfigurations (one for each node). Both \concerto~\cite{chardet:hal-02535077,chardet:hal-02737859} and \concertod~\cite{PhilippeOCPR2024:SANER,omond:hal-04372320,omond:hal-04372340} have been the subject of experimental studies to validate the approach and compare it to related work~\cite{dicosmo2014ic,Sokolowskietal}. In particular, and while this is not the main subject of this paper, \concerto and \concertod allow better parallel execution of reconfiguration actions compared to the related work, thus leading to faster reconfigurations.  

To support the development of this decentralized semantics, it appeared necessary to go beyond a pen-and-paper approach and to provide a mechanized formalization of the semantics. To this end, we used Maude~\cite{clavel2024maude,olveczky2017distributed}, a language based on rewriting logic, and consequently perfectly suited to describing a concurrent model. Developing and formalizing the semantics of \concertod in Maude helped us manage the complexity of the model, and clarify it where needed, but it also allowed us to generate tools such as an interpreter and a model-checker for these semantics. 
A Maude program describes a logical theory, while a Maude computation consists of logical deductions using the axioms specified in the theory.



Throughout the paper, we will use an example taken from a real case study from the literature: a wildlife monitoring system that utilizes sensors to capture animal sounds~\cite{LostanlenBBBFL2021:AM} illustrated in Figure~\ref{fig:fullcps}. These sensors are linked to a gateway and calibrated to specific sound frequencies. Reconfiguration is regularly required in this system to adjust the listening frequency of each sensor. However, during the reconfiguration process, the sensor must temporarily cease listening. Similarly, if the gateway fails to receive data from a sensor, the sensor is required to halt its observation activities. The collected data is stored in a remote database. 
In such a use case, disconnections are common between listeners and sensors. A central coordination of the reconfiguration program is typically a single point of failure, which makes the system fully inactive during unavailability. Furthermore, when facing disconnections, it is more difficult to maintain the global state of the system, which slows down the process: synchronization with the central entity is required even for purely local actions. A decentralized language such as \concertod is preferable in such a context.

\input{figures/cps_full_overview}

In this paper, we give an overview of the semantics of \concertod through its Maude formalization. The complete Maude specification with all the rules is available at \url{https://doi.org/10.5281/zenodo.12786127}.

First, Section~\ref{sec:components} presents how the life cycles of components, and the creation and modification of a components assembly are formalized in \concertod. Second, Section~\ref{sec:semantics} gives an overview of the main semantics rules of \concertod and how are formalized communications in \concertod. Finally, Section~\ref{sec:RelatedWork} and Section~\ref{sec:Conclusion} respectively give the related work and a conclusion on this contribution.

%% file: figures/cps_full_overview.tex
\begin{figure}[t]
\centering
\begin{tikzpicture}
\node[draw] (database) {database};
\node[draw, right=2cm of database] (system) {system};
\node[draw, right=2cm of system] (listener1) {listener$_1$};
\node[draw, right=3cm of listener1] (sensor1) {sensor$_1$};
\node[draw, below=0.6cm of listener1] (listenern) {listener$_n$};
\node[draw, right=3cm of listenern] (sensorn) {sensor$_n$};

\node[below=0.18cm of listener1] (dots1) {...};
\node[below=0.2cm of sensor1] (dots2) {...};

\provideusetwolabels{database}{system}{db}{db\_service}
\provideusetwolabels{system}{listener1}{sys}{sys\_service}
\provideusesinglelabel{listener1}{sensor1}{2}
\provideusesinglelabel{listenern}{sensorn}{2}

\node [draw, dash pattern=on 1pt off 1pt, line width=0.5pt, line width=0.5pt, fit=(database), inner sep=4pt] (box1) {};
\node [below=0.001cm of box1, yshift=0.1cm, xshift=0.7cm] {\tiny node$_1$};

\node [draw, dash pattern=on 1pt off 1pt, line width=0.5pt, line width=0.5pt, fit=(system) (listener1) (listenern), inner sep=4pt] (box2) {};
\node [below=0.001cm of box2, yshift=0.1cm, xshift=2.4cm] {\tiny node$_2$};

\node [draw, dash pattern=on 1pt off 1pt, line width=0.5pt, line width=0.5pt, fit=(sensor1), inner sep=4pt] (box3) {};
\node [below=0.001cm of box3, yshift=0.1cm, xshift=0.7cm] {\tiny node$_{2+1}$};

\node [draw, dash pattern=on 1pt off 1pt, line width=0.5pt, fit=(sensorn), inner sep=4pt] (box4) {};
\node [below=0.001cm of box4, yshift=0.1cm, xshift=0.7cm] {\tiny node$_{2+n}$};

\end{tikzpicture}

\caption{Full overview of the CPS use case with $2+n$ nodes hosting \emph{database}, \emph{system}, $n$ \emph{listeners}, and $n$ \emph{sensors}.}
\label{fig:fullcps}
\end{figure}
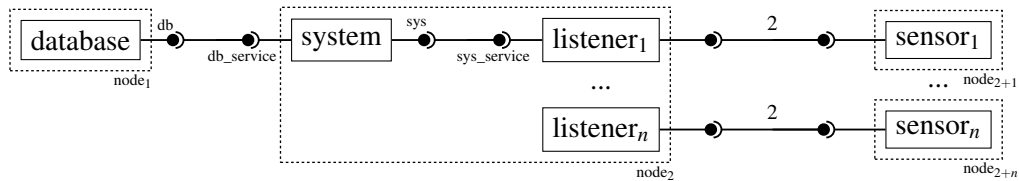

%% file: sections/components.tex
This section presents the structural concepts of \concertod as well as its reconfiguration language. Some elements of the formalization in Maude are included, as well as examples based on our case study.

\subsection{Control components}


In \concertod, a distributed software system is modeled as an assembly of control component instances, linked together via \emph{ports} (\ie interfaces), to represent dependencies in their life cycles (\eg data exchanges, or other interactions between components). There are two types of ports in such connections: \emph{provide} ports and \emph{use} ports. 
A provide port represents information or a resource offered by a component when the port is active. Conversely, a use port indicates that a component requires some information or resources to perform. 

Each piece of software (\ie component, service), identified by a unique identifier, is an instance of a control component type, that models a component's life cycle as a set of \emph{places} and \emph{transitions}. Each port is bound to a subset of the life cycle (\ie places and transitions) that is called a group. 
Places represent milestones of the life cycle. A specific initial place serves as the starting point for the component's life cycle. Transitions represent concrete actions to perform between places (\eg admin commands).

The dynamic nature of components is captured by their \emph{behaviors}. A behavior is a subset of the component's transitions. At any given point, a control component instance executes one behavior, \ie only the transitions in this behavior can be fired. Requests to change the active behavior for a component (see reconfiguration actions in Section~\ref{ssec:language}) are queued and executed in the order in which they are received: a behavior remains active until no more transitions in it can be fired, at which point the behavior is changed to the next requested behavior, if any. This mechanism makes up the behavioral interface of components.


\bigskip\noindent\textbf{\textit{Informal example. }}
Figure~\ref{fig:fullcps} depicts an assembly modeling our use case, where distributed sensors interact with a system through listeners. Physically, the components are distributed on several nodes. The first node serves as a host for the \emph{database} component responsible for storing recorded data obtained from sensors. The second node comprises the components \lstinline[language=maude,basicstyle=\ttfamily]!system! and \lstinline[language=maude,basicstyle=\ttfamily]!listener!. The \emph{system} component establishes a connection with the database through a \emph{use} port (represented using the UML notation), enabling the usage of the database's service. Furthermore, the \emph{system} is interconnected with $n$ \emph{listeners}, each corresponding to a program responsible for monitoring and reconfiguring remote sensors. These listeners are connected to \lstinline[language=maude,basicstyle=\ttfamily]!system! via their use port. Finally, each \lstinline[language=maude,basicstyle=\ttfamily]!sensor!$_i$ is hosted on a node and is connected to its associated \lstinline[language=maude,basicstyle=\ttfamily]!listener!$_i$.

Figure~\ref{fig:listenersensorreconf} shows the internals of some components of this assembly, omitting others (\ie \emph{database}, \emph{system}) for clarity. Each listener has four places: \lstinline[language=maude,basicstyle=\ttfamily]!off!, \lstinline[language=maude,basicstyle=\ttfamily]!paused!, \lstinline[language=maude,basicstyle=\ttfamily]!configured!, and \lstinline[language=maude,basicstyle=\ttfamily]!running!; and three behaviors \lstinline[language=maude,basicstyle=\ttfamily]!deploy!, \lstinline[language=maude,basicstyle=\ttfamily]!update!, and \lstinline[language=maude,basicstyle=\ttfamily]!destroy!. The sensors have five places: \lstinline[language=maude,basicstyle=\ttfamily]!off!, \lstinline[language=maude,basicstyle=\ttfamily]!provisioned!, \lstinline[language=maude,basicstyle=\ttfamily]!installed!, \lstinline[language=maude,basicstyle=\ttfamily]!configured!, and \lstinline[language=maude,basicstyle=\ttfamily]!running!; and also three behaviors: \lstinline[language=maude,basicstyle=\ttfamily]!start!, \lstinline[language=maude,basicstyle=\ttfamily]!pause!, \lstinline[language=maude,basicstyle=\ttfamily]!stop!. Each listener has two connections with its respective sensor. Through the first connection, the listener exposes the configuration to apply on the sensor (\eg frequency of listening). Through the second connection, the listener offers a service to which data can be sent by the sensor when observing.

\bigskip\noindent\textbf{\textit{Maude specification. }}
Let us present the definitions of sorts, subsorts, and operators used to encode the above syntactic aspects of the \concertod model, starting from elementary concepts to the construction of a net of \concertod components. In Maude, a type hierarchy can be defined using the keywords \lstinline[language=maude,basicstyle=\ttfamily]!sort! and \lstinline[language=maude,basicstyle=\ttfamily]!subsort!. An operator definition starts with the keyword \lstinline[language=maude,basicstyle=\ttfamily]!op! followed by the operator name and type signature; several operators with the same signature can be defined using \lstinline[language=maude,basicstyle=\ttfamily]!ops!.

A component type is defined by a set of places, an initial place, and a set of transitions. The definition also includes the behaviors (sets of transitions) of the component type, and its use and provide ports, each bound to a group of places. In the definition below, some details are omitted for simplicity.

We use predefined data structures and the module system of Maude to import the parameterized module \lstinline[language=maude,basicstyle=\ttfamily]!SET{Place}! and define the sort \lstinline[language=maude,basicstyle=\ttfamily]!Places!  to be a supersort of \lstinline[language=maude,basicstyle=\ttfamily]!SET{Place}!. Generally, given a sort \lstinline[language=maude,basicstyle=\ttfamily]!T!, we let \lstinline[language=maude,basicstyle=\ttfamily]!Ts! stand for \lstinline[language=maude,basicstyle=\ttfamily]!SET{T}! and \lstinline[language=maude,basicstyle=\ttfamily]!QT! stand for \lstinline[language=maude,basicstyle=\ttfamily]!LIST{T}!.

\begin{lstlisting}[language=maude, label=ComponentType ]
sorts Place InitialPlace Transition Behavior UsePort ProvidePort GroupUse GroupProvide 
      ComponentType .
sorts Places Transitions Behaviors GroupUses GroupProvides .
subsort InitialPlace < Place .

--- [...]
op b(_) : Transitions -> Behavior [ctor] .
op g(_?_) : UsePort Places -> GroupUse [ctor] . 
op g(_!_) : ProvidePort Places -> GroupProvide [ctor] .  

op { places: _, 
     initial: _, 
     --- [...]
     transitions: _, 
     behaviors: _, 
     groupUses: _,
     groupProvides: _ 
   } : 
   Places InitialPlace 
   --- [...]
   Transitions Behaviors GroupUses GroupProvides -> 
   ComponentType 
  [ctor] .
\end{lstlisting}

\bigskip\noindent\textbf{\textit{Example in Maude. }}
We can now describe the component type \lstinline[language=maude,basicstyle=\ttfamily]!sensor! (an instance of which is displayed on the right of Figure~\ref{fig:listenersensorreconf}). We first define a few constants corresponding to the element of the component type, then the type itself:

\begin{lstlisting}[language=maude, label=ComponentTypeExample]
ops Running Configured Installed Provisioned Off : -> Place .
op Off : -> InitPlace .
--- [...]
ops RcvService ConfigService : -> UsePort .
ops Start11 Start12 Start13 Start2 Start3 Start4 Pause1 Stop1 : -> Transition .
ops Start Pause Stop : -> Behavior .

eq Deploy = b(Start11,Start12,Start13,Start2,Start3,Start4) .
eq Pause = b(Pause1) .
eq Stop = b(Stop1) .
op sensor : -> ComponentType .
eq sensor = 
  { places: Running, Configured, Installed, Provisioned, Off, 
    initial: Off, 
    --- [...] 
    transitions: (Start11,Start12,Start13,Start2,Start3,Start4,Pause1,Stop1), 
    behaviors: (Start, Pause, Stop), 
    groupUses: g(RcvService \lstinline[language=maude,basicstyle=\ttfamily]! (Configured, Running)), 
                g(ConfigService \lstinline[language=maude,basicstyle=\ttfamily]! (Configured, Installed, Running)), 
    groupProvides: empty } .
\end{lstlisting}

\subsection{Reconfiguration language and assembly of components}
\label{ssec:language}

To allow the modification of assemblies, \concertod proposes a simple imperative language for writing reconfiguration programs, that offers four commonly used topological actions to create or modify the existing assemblies: 
$add(id_c, c)$, $del(id_c)$ (creation and deletion of control component instances), 
$con(id_{c1}, u, id_{c2}, p)$,  $dcon(id_{c1}, u, id_{c2}, p)$ (connection and disconnection between two control component instances), where $id_c$ is an instance identifier, $c$ a component type, $u$ a use port, $p$ a provide port.
Besides these actions, two additional actions manage the execution of behaviors: 
$pushB(id_c, b, id_b)$ to request the execution of a behavior $b$ on the component instance $id_c$; and 
$wait(id_c, id_b)$ to synchronize onto the end of a given behavior. $id_c$ is a component instance identifier, and $id_b$ the identifier of a given $pushB$.

As a decentralized coordination model, \concertod considers a system of $n$ nodes, and a partition of component instances over these nodes. Each node operates its own local \concertod controller and runs its own reconfiguration program. Concrete communications allow the synchronization of actions between paired components. This is an evolution of previous work on the \concerto model~\cite{CHARDET2021102582}, in which a single central entity executed the reconfiguration and synchronization of components, and communications were implicit.
Two examples of \concertod reconfiguration programs in our case study are given in Listings~\ref{lst:nodeListeners} and~\ref{lst:site:i}. 


\begin{figure}[h]
\begin{minipage}[t]{0.49\columnwidth}
  \begin{lstlisting}[label=lst:nodeListeners,caption=Listeners reconfiguration on $\mathit{node}_2$,
  language=concertod]
for i in range(nb_listener):
  pushB(listener$_\text{i}$, update, 2+i*2)
  pushB(listener$_\text{i}$, deploy, 3+i*2)
  \end{lstlisting}
\end{minipage}
\hfill
\begin{minipage}[t]{0.49\columnwidth}
  \begin{lstlisting}[label=lst:site:i,caption=One sensor reconfiguration on $\mathit{node}_{2+i}$,language=concertod]
pushB(sensor$_\text{i}$, pause, 0)
wait(listener$_\text{i}$, 2+i*2)
pushB(sensor$_\text{i}$, start,1)
  \end{lstlisting}
\end{minipage}
\end{figure}

\medskip
In \concertod reconfiguration, programs apply changes to the current assembly of components, \ie component instances and their connection, and to the queue of requested behaviors for each component instance. A specific instance of a component (and its state at any given point of the execution) is specified by its component type, a queue of identified behaviors to be executed by the instance, and a marking that indicates the places reached and transitions fired.
\begin{lstlisting}[language=maude, label= Instance ]
sorts Instance Id PushedBehavior TransitionEnding Marking .

op (_;_) : Id Behavior -> PushedBehavior [ctor] .
op m(_,_,_) : Places Transitions TransitionEndings -> Marking [ctor] .
op  {   type: _, 
        queue: _, 
        marking: _ 
    } : 
    ComponentType List{PushedBehavior} Marking -> Instance [ctor] .
\end{lstlisting}

To illustrate this, the definition below describes an instance  \lstinline[language=maude,basicstyle=\ttfamily]!sensor1! of type \lstinline[language=maude,basicstyle=\ttfamily]!sensor!, in a state where only the place \lstinline[language=maude,basicstyle=\ttfamily]!running! is marked, and a single behavior \lstinline[language=maude,basicstyle=\ttfamily]!pause! is pending in the queue of behaviors. This corresponds to the state (0) in Figure~\ref{fig:listenersensorreconf}.

\begin{lstlisting}[language=maude]
eq instanceS1 = {   type: sensor,
                    queue: (0 ; Pause),
                    marking: m(Running, empty, empty) } .
\end{lstlisting}

In order to describe an assembly, it is also necessary to specify the connections between the ports of its component instances, through their identifiers.

\begin{lstlisting}[language=maude, label= Connection ]
sort Connection .
sorts Connections .
op (_,_)--(_,_) : Id UsePort Id ProvidePort -> Connection [ctor] .
\end{lstlisting}

Here is an example of such connections between \lstinline[language=maude,basicstyle=\ttfamily]!node2! and \lstinline[language=maude,basicstyle=\ttfamily]!node3!:
\begin{lstlisting}[language=maude]
eq connectionSL1 = (sensor1, RcvService)--(listener1, Rcv) . 
eq connectionSL2 = (sensor1, ConfigService)--(listener1, Config) .
\end{lstlisting}

We now define actions that can be executed to perform a reconfiguration on a \concertod assembly.

\begin{lstlisting}[language=maude, label = Program]
sorts Action Program .
subsort List{Action} < Program .

op add(_,_) : Id ComponentType -> Action [ctor] .
op del(_) : Id -> Action [ctor] .
op pushB(_,_,_) : Id Behavior Id -> Action [ctor] .
op con(_) : Connection -> Action [ctor] .
op dcon(_) : Connection -> Action [ctor] .
op wait(_,_) : Id Id -> Action [ctor] .
\end{lstlisting}

The reconfiguration program given in Listing~\ref{lst:site:i} (as instantiated specifically for \lstinline[language=maude,basicstyle=\ttfamily]!node3!) can thus be specified in Maude as follows:

\begin{lstlisting}[language=maude,basicstyle=\ttfamily,numbers=none]
eq programNode3 = 
pushB(sensor1, start, 0) wait(listener1, 4) pushB(sensor1, start, 1) . 
\end{lstlisting}



%% file: sections/full_example.tex
\begin{figure}[t]
\centering
\includegraphics[scale=0.8]{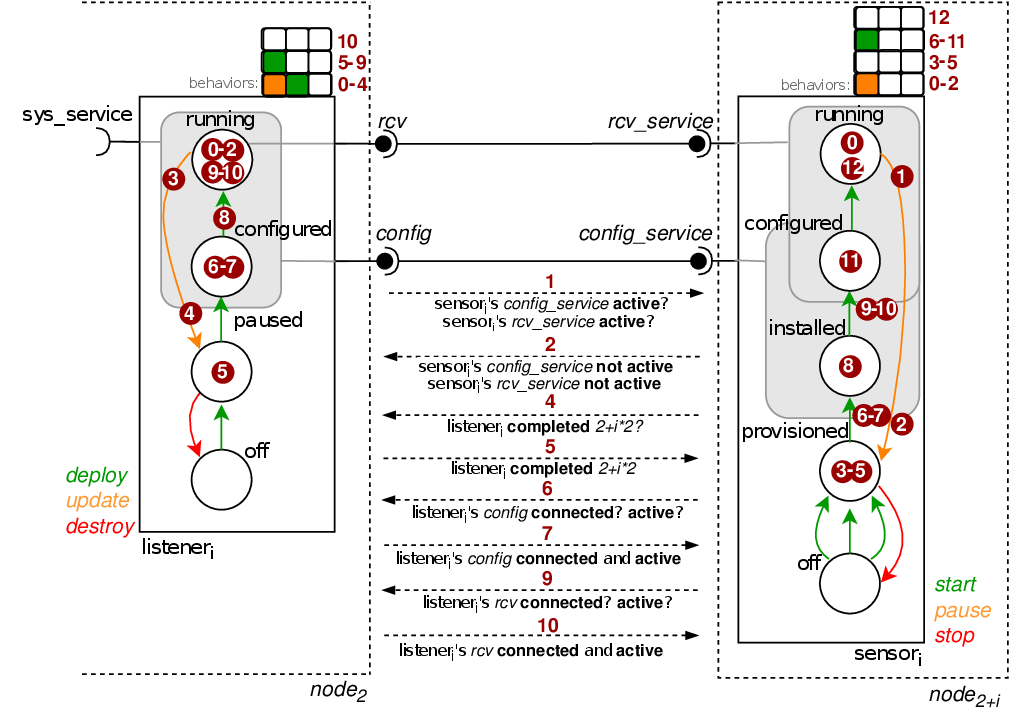}
\caption{Control components of \texttt{listener$_\text{i}$} on $\mathit{node}_2$ and
\texttt{sensor$_\text{i}$} from $\mathit{node}_{2+i}$. State example when applying the reconfiguration plans of Listings~\ref{lst:nodeListeners} and ~\ref{lst:site:i}.}
\label{fig:listenersensorreconf}
\end{figure}


\concertod is equipped with a small-step semantics. We first illustrate the execution of a reconfiguration on our use case to give an intuition of this semantics. 

\medskip
We consider deployed and running components, \ie the places \texttt{running} are marked in all listener and sensor components. From there, we aim to trigger an update of each sensor's listening frequency. Hence, each listener has to pause to change its configuration, forcing the sensors to pause as well. 

\medskip
Listings~\ref{lst:nodeListeners} and~\ref{lst:site:i} give the reconfiguration programs respectively for all listeners (all hosted on the same node $\mathit{node}_2$) and one sensor $i$ (corresponding to the update of the frequency of each sensor). These programs are executed concurrently on each node. The execution of these scripts is illustrated in Figure~\ref{fig:listenersensorreconf} to give the reader an intuition of the semantics of \concertod. Some possible steps of the execution are represented by a number representing the current configuration of the system (\ie a snapshot). 
Using these numbers, three pieces of information are given: 
(i) the marking, represented by the red dots on marked places and transitions; (ii) the status of the behavior queue, and (iii) the coordination information required between nodes introduced by decentralized execution of \concertod. 
In the following, we describe the state of each configuration according to its number. We decompose each state, and we highlight communication steps using the notation $\Delta$.

\begin{enumerate}
\setlength{\itemsep}{0pt}
\setcounter{enumi}{-1}

\item \label{exec0} Both $\mathit{listener}_i$ and $\mathit{sensor}_i$ services are running. Thus, the \texttt{running} places are marked. The behaviors to trigger the update are pushed in the queue, \ie \texttt{update} and \texttt{deploy} for $\mathit{listener}_i$, and \texttt{pause} for $\mathit{sensor}_i$.
The behavior \texttt{start} for $\mathit{sensor}_i$ is not pushed due to wait action of a behavior of $\mathit{listener}_i$.

\item By triggering \texttt{pause},  $\mathit{sensor}_i$ leaves the place \texttt{running}. 

\item[$\Delta$]  $\mathit{listener}_i$ needs to know if any component is using its ports before executing \texttt{update}. Then, $\mathit{listener}_i$ requests information on $\mathit{sensor}_i$'s use ports $\mathit{rcv\_service}$ and $\mathit{config\_service}$, to know if they are active or not.

\item \label{exec2} $\mathit{sensor}_i$ ends its previously fired transition.
\item[$\Delta$] $\mathit{sensor}_i$ answers $\mathit{listener}_i$ about its use ports $\mathit{rcv\_service}$ and $\mathit{config\_service}$, indicating that both are inactive. 

\item \label{exec3} $\mathit{listener}_i$ can begin the \texttt{update} behavior, therefore deactivating its provide ports. 
$\mathit{sensor}_i$ removes the \texttt{pause} behavior from its queue, as no more transitions in this behavior can be fired. 
\item[$\Delta$] The information sent in~\ref{exec2} is received by $\mathit{listener}_i$, allowing it to start its \texttt{update}.

\item $\mathit{listener}_i$ finishes its transition related to its \texttt{update} behavior. 
\item[$\Delta$] $\mathit{sensor}_i$ is now waiting the behavior identified by 2+i*2, and runs on $\mathit{listener}_i$, to be ended. Then it sends a request to $\mathit{listener}_i$ to know if the behavior 2+i*2 (\texttt{update}) is completed. 

\item The behavior \texttt{update} of $\mathit{listener}_i$ is retrieved from its queue.
\item[$\Delta$] $\mathit{sensor}_i$ is informed that this behavior is completed, which allows it to continue its execution, i.e., to push and execute the transitions of the \texttt{start} behavior.

\item After pushing the behavior \texttt{start} and completing its first transition.
\item[$\Delta$] $\mathit{sensor}_i$ sends a request to determine whether the port $\mathit{config}$ of $\mathit{listener}_i$ is connected and active, before entering in $\mathit{config\_service}$ group.

\item $\Delta$ $\mathit{sensor}_i$ receives information on $\mathit{listener}_i$'s port $\mathit{config}$. It is now active and connected.

\item $\mathit{sensor}_i$ enters the place \texttt{installed}. $\mathit{listener}_i$ begins the activation of its service.

\item \label{exec9} $\mathit{listener}_i$ activates its service and provides the port $\mathit{rcv}$. 
\item[$\Delta$] $\mathit{sensor}_i$ sends a request asking whether the port $\mathit{rcv}$ is active and connected before entering the place \texttt{configured}.

\item  $\mathit{listener}_i$ removes the \texttt{deploy} behavior from its queue. 
\item[$\Delta$] $\mathit{sensor}_i$ get the information that $\mathit{listener}_i$ provides the port $\mathit{rcv}$.

\item $\mathit{sensor}_i$  enters the place \texttt{configured}. 

\item $\mathit{sensor}_i$ reactivates its service and removes the \texttt{start} behavior from its queue.

\end{enumerate}

The rest of this section gives some formal elements of the small-step semantics of \concertod. Because of space reasons, we cannot fully detail all rules of the semantics. Instead, the section first illustrates one execution rule at a reconfiguration program level, one execution rule at a component level, and then details communications, \ie the main formalization challenge in a decentralized model as \concertod. 

%% file: sections/semantics.tex
Let us first describe the semantics rules that govern the execution of actions in reconfiguration programs. Recall that each node executes its own reconfiguration program, affecting components hosted on that node. Actions to add and delete component instances are straightforward. Actions for connections and disconnections of components are also fairly standard. They only modify the view that one node has of the component's topology. When components to (dis)connect are located on different nodes, both nodes should eventually execute their own similar (dis)connection actions, however, the node-local views of the topology do not have to be kept continuously synchronized. Finally, we turn to the \lstinline[language=maude,basicstyle=\ttfamily]!pushB! action, which is specific to the \concertod model. The rule in Listing~\ref{PushingBehaviorRule} gives the semantic of that action, describing the modification of a node (as defined in Section~\ref{sec:communications}). The action adds a behavior request to the queue of behaviors of one component. The target component is indicated by its identifier, and the behavior type is indicated, as well as an identifier. The latter is used for synchronization points, in order to distinguish between multiple executions of the same behavior type.


The behavior requests determine the active behavior of a component instance at a given point, which plays a role in the evolution of this component. We now describe the rules of this evolution.
As described in Section~\ref{sec:components}, the life cycles of components are modeled by places (milestones in the reconfiguration) and transitions between places (reconfiguration actions). A marking on places and transitions indicates the current state of the reconfiguration. When a place is marked, its outgoing transitions can be fired. The place is then unmarked, and the fired transitions are marked. Conversely, when all the incoming transitions toward a place are finished, those transitions are unmarked and the place is entered. This model is well suited to represent concurrent execution, in particular, multiple transitions can be marked simultaneously, corresponding to parallel execution of reconfiguration actions.

\begin{lstlisting}[language=maude, caption={The rule to execute the \texttt{pushB} action. The rule applies to a node (as defined in Section~\ref{sec:communications}) but rewrites a single component instance in that node.},label = PushingBehaviorRule]
crl [PushingBehavior] : 
 < --- [...]
   instances: (Id1 |-> { type: Ct, queue: Qb, marking: M }) , Isx,
   --- [...] 
   program: pushB(Id1, B, IdB) RPx,
   --- [...] >  
  =>   
 < --- [...]
   instances: ( Id1 |-> { type: Ct, queue: append(Qb,(IdB ; B)), marking: M } ), Isx,
   --- [...]
   program: RPx,
   --- [...]
 >   
  if (B in (Ct).behaviors) .
\end{lstlisting}


This evolution of markings is desribed by four rules: \emph{FiringTransitions}, \emph{EndingTransition}, \emph{EnteringPlace}, and \emph{FinishingBehavior}.
For space reasons, we only detail one of these rules: \emph{Firing Transitions} given in Listing~\ref{FiringTransitionRule}. It modifies the marking in one component on one node, namely unmarking a place \lstinline[language=maude,basicstyle=\ttfamily]!P! and marking the outgoing transitions of \lstinline[language=maude,basicstyle=\ttfamily]!Ts! (only those transitions that belong to the active behavior of the component). Note that this is a conditional rule, as indicated by the keyword \lstinline[language=maude,basicstyle=\ttfamily]!crl! and the boolean conditions after the rule. The first condition merely ensures that there are transitions to fire. The second condition relies on the predicate \lstinline[language=maude,basicstyle=\ttfamily]!isSafeToFire! to check dependencies towards other components, modeled by ports, are not violated by the firing of the transitions. In particular, this predicate is true only if the firing of the transitions does not lead to deactivating a provide port that is being used, \ie connected to an active use port. The rule to end a transition (not given) is somewhat similar but instead requires that use ports attached to the reached place are connected to an active provide port. Thus, ports impose inter-component synchronization conditions on the execution.

\begin{lstlisting}[language=maude, caption={The rule to fire a transition. As before, this rule modifies a component instance in a node.},label = FiringTransitionRule]
crl [FiringTransitions] :
    < nodeInventory: Idsx,
      instances: (Id1 |->   {   type: Ct, 
                                queue: (IdB ; b(TsB)) Qb, 
                                marking: m((P, Ps),Ts,Tes) 
                            }) , Isx,
      connections: Csx,
      --- [...]> 
    =>
    < nodeInventory: Idsx,
      instances:  (Id1 |->  {   type: Ct, 
                                queue: (IdB ; b(TsB)) Qb, 
                                marking: m(Ps,union(Ts, getTransitionsofPlace(TsB,P) ),Tes) 
                            }), Isx , 
      connections: Csx,
      --- [...]> 
    if (    getTransitionsofPlace(TsB,P) =/= empty and
            isSafeToFire(   Idsx,
                            m(Ps,union(Ts, getTransitionsofPlace(TsB,P) ),Tes) ,
                            (Id1 |->    {   type: Ct,
                                            queue: (IdB ; b(TsB)) Qb, 
                                            marking: m(P, Ps,Ts,Tes)
                                        } ,Isx),
                            eStatex,
                            getPConnectionsofCInst(Id1, Csx))) .
        ) .
\end{lstlisting}




%% file: sections/communications.tex

Firing or ending a transition on one component may require checking the activity status of a port on another component. In previous work~\cite{CHARDET2021102582}, the coordination was assumed to be carried out by a central entity that kept track of the status of every port. Instead, \concertod is meant to represent a decentralized process, it is therefore necessary to explicitly model communications between the nodes that host components.

The following boolean information may be needed for synchronization, and can be called on remote nodes: 
(1) the completion of a $dcon$ action; 
(2) the completion of a behavior (\ie $wait$ action); 
(3) the activity of a use or provide port. 

When evaluating the status of one of the above queries (such as the status of a port in the predicate \lstinline[language=maude,basicstyle=\ttfamily]!safeToFire!), we distinguish the case where the element of interest ($dcon$ action, behavior, port etc.) belongs to the local node, from the case where messages must be exchanged between nodes:
\begin{lstlisting}[language=maude]
  eq question(R1,Ids,RcvAns, Is,Cs,RP ) =   if (isLocal(R1,Ids)) then 
                                                localQuestion(R1,Is,Cs,RP) 
                                            else externQuestion(R1,RcvAns) fi .
\end{lstlisting}

Essentially, each \concertod node maintains a localized perspective of its components and communicates with neighboring nodes (\ie other \concertod controllers hosting connected control components). In \concertod, an asynchronous message-based communication model, increasingly favored in distributed systems, manages the communication~\cite{omond:hal-04372320}.
To achieve asynchronous communications each node is equipped with an incoming queue of messages. Messages must transit through this queue before being effectively received (this assumes that the order of messages is preserved). 

A message can either be a \lstinline[language=maude,basicstyle=\ttfamily]!Question! or an \lstinline[language=maude,basicstyle=\ttfamily]!Answer!. The former is aimed at a specified component instance and contains one of several possible queries while an \texttt{Answer} gives the boolean value corresponding to a question. It can take the values {true, false}.
\begin{lstlisting}[language=maude, caption={Constructors for Question, Answer and Message}, label = Request]
sorts Query .
op isActive(_) :  Port -> Query [ctor] .  
op isRefusing(_) : Port -> Query [ctor] .    
op isConnected(_) : Connection -> Query [ctor] .
op isCompleted(_) : Id -> Query [ctor] .
op onDisconnect(_) : Connection -> Query [ctor] . 
op [ dst: _ , query: _ ] : Id Query -> Question [ctor] .
op [ question: _, value: _ ] : Question Bool -> Answer [ctor] .
op mkMsg : Question -> Message [ctor] .
op mkMsg : Answer   -> Message [ctor] .
\end{lstlisting}



For example, the question \lstinline[language=maude,basicstyle=\ttfamily]![ dst: listener1, query: isActive(Rcv) ]! can be sent by node3 (which hosts \lstinline[language=maude,basicstyle=\ttfamily]!sensor1!) to node2 (which hosts \lstinline[language=maude,basicstyle=\ttfamily]!listener1!) to check whether the port \lstinline[language=maude,basicstyle=\ttfamily]!Rcv! of \lstinline[language=maude,basicstyle=\ttfamily]!listener1! is active. This corresponds to the state ~\ref{exec9} in Figure~\ref{fig:listenersensorreconf}. The answer \lstinline[language=maude,basicstyle=\ttfamily]![ req: [ dst: listener1, query: isActive(Rcv) ], value: true ]! is the answer which will be returned at state (10) to indicate that the port is indeed active. For more details about the remaining queries, the reader can refer to~\cite{CHARDET2021102582}.

\medskip
A node is specified by the set of identifiers of all the component instances involved in the reconfiguration programs, a mapping (called an inventory) that associates these identifiers to their associated nodes, the connections between local component instances and external instances, and the local reconfiguration program to be executed on the node.
A node is additionally specified by five parameters used for communication: (1) a local vision of the state of external components \lstinline[language=maude,basicstyle=\ttfamily]!externState!; (2) a queue of pending questions waiting for an answer \lstinline[language=maude,basicstyle=\ttfamily]!pendingQuestions!; (3) a queue of outgoing questions asked to other nodes \lstinline[language=maude,basicstyle=\ttfamily]!outgoingQuestions!; (4) a queue of incoming messages \lstinline[language=maude,basicstyle=\ttfamily]!incomingMsgs! including incoming questions asked by other nodes and answers received from other nodes to previously sent questions; and finally (5) a history of previously sent questions (to avoid redundant messages for already pending requests).
%
Received answers are stored in \lstinline[language=maude,basicstyle=\ttfamily]!externState! as a mapping that associates requests to the value of the answer. This represents the local node's vision of the status of other components. This information is updated each time external information is needed in a semantics rule. 
%


\begin{lstlisting}[language=maude, caption={Local configuration}, label = LocalConfiguration ]
 op  <  nodeInventory: _, 
        instances: _, 
        connections: _,
        program: _,
        externState: _,
        pendingQuestions: _, 
        outgoingQuestions: _, 
        history: _,
        incomingMsgs: _ 
      > : 
      Ids Instances Connections Program Map{Question, Bool} List{Question} List{Question} Set{Question} List{Message} 
      -> LocalConfiguration 
\end{lstlisting}  
For example, the description of the local configuration of \texttt{node3} in state~\ref{exec0} is as follows:
\begin{lstlisting}[language=maude, label = LocalConfigurationE]
eq ConfNode3 =   
     < nodeInventory: sensor1, 
       instances: { sensor1 |-> instanceS1 }, 
       connections: { connectionSL1, connectionSL2 }, 
       program: programNode3, 
       externState: eStateNode3, 
       pendingQuestions: nil, 
       outgoingQuestions: nil, 
       history: empty,
       incomingMsgs: nil >
\end{lstlisting}  
where \lstinline[language=maude,basicstyle=\ttfamily]!instanceS1!, \lstinline[language=maude,basicstyle=\ttfamily]!connectionSL1!, \lstinline[language=maude,basicstyle=\ttfamily]!connectionSL2! and \lstinline[language=maude,basicstyle=\ttfamily]!programNode3! are the elements previously described for our use case in Sections~\ref{sec:components} and \ref{sec:communications}. \lstinline[language=maude,basicstyle=\ttfamily]!sensor1! is the identifier of \lstinline[language=maude,basicstyle=\ttfamily]!instanceS1! and \lstinline[language=maude,basicstyle=\ttfamily]!eStateNode3! is the external state (\ie mapping of questions and received answers sent by \texttt{node3}). Following the deployment steps that preceded the reconfiguration state~\ref{exec0}, \lstinline[language=maude,basicstyle=\ttfamily]!eStateNode3! of \texttt{node3} is described as follows:
\begin{lstlisting}[language=maude, label = ReceivedAnswers]
eq eStateNode3 =  
    [ dst: listener1, query: isRefusing(Rcv) ] |-> false , 
    [ dst: listener1, query: isConnected(connectionSL1) ] |-> true , 
    [ dst: listener1, query: isRefusing(Config) ] |-> false , 
    [ dst: listener1, query: isConnected(connectionSL2) ] |-> true . 
\end{lstlisting} 



The rest of this section gives the execution semantics of communication in \concertod using Maude rewrite rules, which operate on a system of local configurations of nodes. 
The associated rules for communication are presented in listings \ref{RequestRule}, \ref{SendEvaluationRule}, \ref{ReceiveRequestRule} and \ref{ReceiveEvaluationRule}. The rules consider two nodes \lstinline[language=maude,basicstyle=\ttfamily]!x! and \lstinline[language=maude,basicstyle=\ttfamily]!y!.

First, Listing~\ref{RequestRule} describes the rule \lstinline[language=maude,basicstyle=\ttfamily]![SendQuestion]! sending of a request \lstinline[language=maude,basicstyle=\ttfamily]![ dst: Dst, query: Q ]! from node \lstinline[language=maude,basicstyle=\ttfamily]!x! to node \lstinline[language=maude,basicstyle=\ttfamily]!y!. The destination node \lstinline[language=maude,basicstyle=\ttfamily]!y! is determined since the identifier of the instance of the request \lstinline[language=maude,basicstyle=\ttfamily]!Dst! appears in the inventory of node \lstinline[language=maude,basicstyle=\ttfamily]!y!. Sending the request implies adding to the incoming messages of \lstinline[language=maude,basicstyle=\ttfamily]!y! the message \lstinline[language=maude,basicstyle=\ttfamily]!mkMsg([ dst: Dst, query: Q ])!. The sent request will also be inserted in the history \lstinline[language=maude,basicstyle=\ttfamily]!Hx! of node \lstinline[language=maude,basicstyle=\ttfamily]!x! to avoid the request being sent multiple times while waiting for an answer.


\begin{lstlisting}[language=maude, caption={The rule to send a question, that rewrites two nodes (unaffected variables are omitted).},label = RequestRule]
rl [SendQuestion] : 
  < --- [...]
    outgoingQuestions: [ dst: Dst, query: QY ] OutQx,
    history: Hx,
    --- [...]
  > ,
  < nodeInventory: (Dst, Idsy),
    ---[...]
    incomingMsgs: iMsgsy > 
  =>
  < --- [...]
    outgoingQuestions: OutQx,
    history: ([ dst: Dst, query: Q ], Hx),
    --- [...]
  > , 
  < nodeInventory: (Dst, Ids),
    --- [...]
    incomingMsgs: append(iMsgsy, mkMsg([ dst: Dst, query: QY ])) > 
\end{lstlisting}

Second, Listing~\ref{SendEvaluationRule} describes the sending of an answer from node \lstinline[language=maude,basicstyle=\ttfamily]!x! to node \lstinline[language=maude,basicstyle=\ttfamily]!y! concerning a question \lstinline[language=maude,basicstyle=\ttfamily]!Q! previously sent by \lstinline[language=maude,basicstyle=\ttfamily]!y!. The destination node \lstinline[language=maude,basicstyle=\ttfamily]!y! is chosen when the concerned request \lstinline[language=maude,basicstyle=\ttfamily]!Q! is in its history of sent requests, and has not yet received in an answer to this request (!not occurs(mkMsg(Q), iMsgsy)!). The value of the answer is computed on node \lstinline[language=maude,basicstyle=\ttfamily]!x! by the function \lstinline[language=maude,basicstyle=\ttfamily]!localQuestion! and sent to node \lstinline[language=maude,basicstyle=\ttfamily]!y! by placing the answer in the incoming messages of node \lstinline[language=maude,basicstyle=\ttfamily]!y!. 


It is important to maintain the consistency of shared information, so if a node $x$ sends information that could allow an external component $y$ to change the status of one of its ports $p$, the previously recorded information about $p$ on $x$ is deleted. For example, when a component on node $y$ asks $x$ if one of its provide port $p$ is active, it means that the associated use port $u$ on $y$ may be activated soon after the answer is sent by $x$.
Consequently, the node $x$ that sends the answer on $p$ will simultaneously reset the information in its external state regarding $u$ on $y$ (of course, it should not yet assume that the use port is active). In Listing \ref{SendEvaluationRule}, this update is carried out by the function \lstinline[language=maude,basicstyle=\ttfamily]!resetState!.

\begin{lstlisting}[language=maude, caption={The (conditional) rule to send an answer},label = SendEvaluationRule]
crl [SendAnswer] : 
  < --- [...]
    instances: Isx, 
    connections: Csx,
    program: RPx,
    externState: eStatex,
    pendingQuestions: Q pendingQx, 
    --- [...]
  > ,
  < --- [...]
    history: (Q, Hy),
    incomingMsgs: iMsgsy >   
  =>
  < --- [...]
    instances: Isx, 
    connections: Csx,
    program: RPx,
    externState:resetState(eStatex,getIdsForReset(Csx,Q,localQuestion(Q,Isx,Csx,RPx))),
    pendingQuestions: pendingQx, 
    --- [...]
  > ,
  < --- [...]
    history: Hy,
    incomingMsgs: append(iMsgsy, mkMsg([    question: Q, 
                                            value: localQuestion(Q,Isx,Csx,RPx) ]))
  > 
 if (not occurs(mkMsg(Q), iMsgsy)) .
\end{lstlisting}

Third, Listing~\ref{ReceiveRequestRule} describes the rule for receiving a question on a component local to \lstinline[language=maude,basicstyle=\ttfamily]!x!. This is started when the identifier \lstinline[language=maude,basicstyle=\ttfamily]!Dst! of the request \lstinline[language=maude,basicstyle=\ttfamily]![dst: Dst, query: Q]! is in the set of identifiers of the components of node \lstinline[language=maude,basicstyle=\ttfamily]!x!. The request is evaluated and the answer is placed in the queue of pending questions.
%

\begin{lstlisting}[language=maude, caption={The rule for receiving a question}, label = ReceiveRequestRule]
rl [ReceiveQuestion] : 
  < nodeInventory: (Dst, Idsx),
    --- [...]
    pendingQuestions: pendingQx,
    --- [...]
    incomingMsgs: mkMsg([dst: Dst, query: QY]) iMsgsx >   
  => 
  < nodeInventory: (Dst, Ids),
   --- [...]
   pendingQuestions: append(pendingQx,[ dst: Dst, query: QY ]),
   incomingMsgs: iMsgsx >   
\end{lstlisting}

Finally, Listing~\ref{ReceiveEvaluationRule} describes the rule for receiving an answer to a question \lstinline[language=maude,basicstyle=\ttfamily]!Q! previously sent by node \lstinline[language=maude,basicstyle=\ttfamily]!x! (as indicated by its presence in the history).
The answered value \lstinline[language=maude,basicstyle=\ttfamily]!val! is recorded (!insert(R, val, RcvA)!) and it replaces previous information about the result of request \lstinline[language=maude,basicstyle=\ttfamily]!Q!, if any.
The request \lstinline[language=maude,basicstyle=\ttfamily]!Q! is removed from the history of node \lstinline[language=maude,basicstyle=\ttfamily]!x!, so that a similar request may be sent again in the future.

\begin{lstlisting}[language=maude, caption={The rule for receiving an answer}, label = ReceiveEvaluationRule]
rl [ReceiveAnswer] : 
 < --- [...]
   externState: eStatex,
   --- [...]
   history: (Q,Hx),
   incomingMsgs: mkMsg([ question: Q, value: Val ]) iMsgsx >   
     =>
 < --- [...]
   externState: insert(Q,Val,eStatex),
   --- [...]
   history: Hx,
   incomingMsgs: iMsgsx > 
    
\end{lstlisting}

\bigskip\noindent\textbf{\textit{Example. }}
To illustrate the communication protocol of the above semantics rules, a subpart of the example of Figure~\ref{fig:listenersensorreconf} is used with one listener and one sensor. In step 4, the \lstinline[language=maude,basicstyle=\ttfamily]!sensor! sends a question to the \lstinline[language=maude,basicstyle=\ttfamily]!listener! regarding the completion of its behavior \lstinline[language=maude,basicstyle=\ttfamily]!idb4!. The question is created following the \lstinline[language=maude,basicstyle=\ttfamily]!wait(listener, idb4)! action in the local reconfiguration program of the sensor, and is placed in \lstinline[language=maude,basicstyle=\ttfamily]!outgoingQuestion!. The question is \lstinline[language=maude,basicstyle=\ttfamily]![dst: listener1, query: isCompleted(idb4)]!.
The following communication rules are applied:
\begin{enumerate}
    \item \lstinline[language=maude,basicstyle=\ttfamily]![SendQuestion]! is applied: the question in \lstinline[language=maude,basicstyle=\ttfamily]!outgoingQuestion! on \lstinline[language=maude,basicstyle=\ttfamily]!sensor! is popped and placed into \lstinline[language=maude,basicstyle=\ttfamily]!history! of \lstinline[language=maude,basicstyle=\ttfamily]!sensor!. The \lstinline[language=maude,basicstyle=\ttfamily]!incomingMsgs! of \lstinline[language=maude,basicstyle=\ttfamily]!listener! appends the question received from \lstinline[language=maude,basicstyle=\ttfamily]!sensor!.
    \item \lstinline[language=maude,basicstyle=\ttfamily]![ReceiveQuestion]! is applied: The node \lstinline[language=maude,basicstyle=\ttfamily]!listener! appends the incoming question to its list \lstinline[language=maude,basicstyle=\ttfamily]!pendingQuestions!. The question is removed from \lstinline[language=maude,basicstyle=\ttfamily]!incomingMsgs!.
    \item \lstinline[language=maude,basicstyle=\ttfamily]![SendAnswer]! is applied: The question in \lstinline[language=maude,basicstyle=\ttfamily]!pendingQuestions! on \lstinline[language=maude,basicstyle=\ttfamily]!listener! is popped. The answer is created and pushed to the \lstinline[language=maude,basicstyle=\ttfamily]!incomingMsgs! of \lstinline[language=maude,basicstyle=\ttfamily]!sensor!. The answer message to the question is \lstinline[language=maude,basicstyle=\ttfamily]![question: [dst: listener1, query: isCompleted(idb4)], value: true]!. The original question is removes from \lstinline[language=maude,basicstyle=\ttfamily]!history! on \lstinline[language=maude,basicstyle=\ttfamily]!sensor!.
    \item \lstinline[language=maude,basicstyle=\ttfamily]![ReceiveAnswer]! is applied: The answer received on \lstinline[language=maude,basicstyle=\ttfamily]!sensor! in \lstinline[language=maude,basicstyle=\ttfamily]!incomingMsgs! is removed and inserted (added or updated) to the \lstinline[language=maude,basicstyle=\ttfamily]!externState! of \lstinline[language=maude,basicstyle=\ttfamily]!sensor! as a mapping \lstinline[language=maude,basicstyle=\ttfamily]![dst: listener, query: isCompleted(idb4)] |-> true]!.
\end{enumerate}
Once completed a pushed behavior cannot be executed again as being uniquely identified. For this reason, no consistency issue can happen in this small example. The \lstinline[language=maude,basicstyle=\ttfamily]!resetState! of the rule \lstinline[language=maude,basicstyle=\ttfamily]!SendAnswer! has no consequences on \lstinline[language=maude,basicstyle=\ttfamily]!externState! in this case.

%% file: RelatedWork.tex
\concerto and \concertod can be considered as component models, such as defined in Component-Based Software Engineering (CBSE). As explained in~\cite{CHARDET2021102582,CHL2023:CSUR} \concerto, and by extension \concertod, differ from usual component models by modeling the life cycle of components instead of modeling the functional code of components. For this reason, both \concerto and \concertod are more comparable to DevOps approaches. Only one other component model can be compared to \concerto: Aeolus~\cite{dicosmo2014ic}. However, Aeolus is more limited than \concerto in terms of parallelism and concurrency. As for \concerto, and unlike \concertod, Aeolus is a centralized model, and it has been formalized manually.

Regarding DevOps approaches for deployments, a few contributions have studied a decentralized approach. In~\cite{6727828,wildetal} each component of the application expresses its dependencies with the other components in a central plan, distributed to the corresponding nodes, deploying their part of the application. Deployment executions are then coordinated between the nodes according to their dependencies. Here, the execution is decentralized as in \concertod. In~\cite{Sokolowskietal}, an extension of the DevOps tool Pulumi is presented to handle both the computation and execution of deployment and update programs in a decentralized manner. However, the three approaches above, and almost all DevOps tools, lack formal specifications: their language is defined by a unique implementation and informal documentation.

To our knowledge, there are only two DevOps contributions that offer formal semantics of their system configuration languages: \smartfrog~\cite{AH2016:JNSM} a tool no longer maintained, and \mupuppet~\cite{FPA2017:ECOOP} a subset of \puppet. 
The semantics of \smartfrog is a denotational semantics of a \emph{compiler} for a core \smartfrog fragment. From the high-level language SF, which handles features such as inheritance, composition, references, etc., the compilation process produces a store, basically a tree of attribute-value pairs. These trees are also the abstractions for system states in \smartfrog's semantics. The authors wrote three implementations (in Scala, Haskell, and OCaml) of the compiler guided by the semantics but not proved correct \wrt to the formal semantics (which is not mechanized). These implementations were randomly tested and a few implementation errors were found. They were also tested against the production compiler: it allowed them to find both a misunderstanding of the semantics of \smartfrog and bugs in the production compiler.
The semantics of \mupuppet is a small-step operational semantics. An implementation of a \mupuppet compiler exists, guided by the semantics but not proved correct \wrt the formal semantics, which is also not mechanized. The output of the compiler is a catalog, \ie a structure close to the stores' output by \smartfrog's compiler. A catalog is also an abstraction for a system state. To help debug \mupuppet manifests, Fu~\cite{FU2019:PHD} also proposed an analysis of provenance~\cite{CCT2009:FTDB} based on the \mupuppet operational semantics. 
In both cases, the formal semantics is not an executable artifact as is our proposal.

Khebbeb \etal also use Maude to formalize adaptation in the Cloud~\cite{KHB2019:IETS} and at the edge~\cite{KHB2020:JSA}. The motivation and approach are however very different from our work. There is no reconfiguration language: the goal is to automatically adapt resources depending on the load of the Cloud system. On the one hand, this work considers the provisioning and de-provisioning of virtual machines, an aspect we do not consider. On the other hand, their concept of service is very simplistic: a service is something that processes requests (and only the number of requests is formalized) and there are no connections between services. Their rewrite rules implement pre-defined elasticity strategies. While they perform model-checking on a small example (3 services, 1 VM, and 2 Fog nodes)~\cite{KHB2020:JSA}, they neither provide any information about the number of states, nor the time required for the verification. In its complexity, our proposal is closer to work that models APIs for \eg~\cite{YCB2024:FASE}. As Yu \etal, to analyze interesting enough case studies, we will need to optimize the model-checking by implementing partial order reduction~\cite{FM2006:WRLA}.


%% file: Conclusion.tex
In this paper, we proposed a formalization in Maude of the decentralized reconfiguration language \concertod, an extension of the centralized reconfiguration language \concerto~\cite{CHARDET2021102582}. In \concertod, the assembly of components is distributed among a set of nodes. Each node is responsible for a local reconfiguration program that may require some coordination with the nodes that host connected component instances. \concertod automatically manages the necessary communications between nodes via asynchronous communications, which involve nodes communicating by exchanging messages through buffers.
The advantage of having a formal semantics with Maude is that it is executable and hence we can check by running examples that the semantics we designed behaves as we expect it to do. 

This is however a first step. We ambition to verify properties of specific assemblies and their reconfiguration programs as well as to verify properties of the semantics itself. 

To verify specific assemblies and their programs, Maude offers on-the-fly model-checking of LTL formulas. Preliminary experiments show that in its current form the semantics does not allows to scale up to the model-checking of realistic applications. We will first explore the application of partial order reduction techniques~\cite{FM2006:WRLA} to limit the state-space explosion problem. Another way to avoid such an explosion is to use symbolic model-checking techniques in particular verification by over-approximation of the set of descendants~\cite{GEN1998:RTA,BCR2019:JCSS}. In these approaches, the goal is to check that a set of bad states does not intersect the over-approximations of the sets of descendants. There are however constraints on how the (approximations of) sets of terms are described (for e.g. regular tree automata) and constraints on the rewriting rules describing the dynamic of the system. 

Verifying properties of the \concertod model itself may require the use of interactive theorem proving (ITP). There are ongoing efforts to develop a dedicated interactive theorem prover for Maude~\cite{nuitp} as well as to translate Maude specifications~\cite{rubio2022theorem} into existing ITP specification languages. The main properties we plan to prove interactively are that the semantics preserves well-formed instances and well-formed systems, as well as that the consistency of the communicated information between nodes is ensured. 

Finally, thanks to \concertod, \concerto could be transformed into a choreographic language (as defined in~\cite{ABB2016:FTPL}). In a choreographic approach \concerto would be the choreography language. A compilation process would then automatically generate the \concertod programs (\ie local projections) on the nodes, and guarantee that required communications between nodes will be performed when required. The formalization may be used to prove that the set of \concertod programs yielded by that process has a behavior that corresponds to the behavior original centralized \concerto program.



%% file: main.bbl
\begin{thebibliography}{10}
\providecommand{\bibitemdeclare}[2]{}
\providecommand{\surnamestart}{}
\providecommand{\surnameend}{}
\providecommand{\urlprefix}{Available at }
\providecommand{\url}[1]{\texttt{#1}}
\providecommand{\href}[2]{\texttt{#2}}
\providecommand{\urlalt}[2]{\href{#1}{#2}}
\providecommand{\doi}[1]{doi:\urlalt{https://doi.org/#1}{#1}}
\providecommand{\eprint}[1]{arXiv:\urlalt{https://arxiv.org/abs/#1}{#1}}
\providecommand{\bibinfo}[2]{#2}

\bibitemdeclare{article}{ABB2016:FTPL}
\bibitem{ABB2016:FTPL}
\bibinfo{author}{Davide~Ancona \surnamestart et~al.\surnameend}
  (\bibinfo{year}{2016}): \emph{\bibinfo{title}{Behavioral Types in Programming
  Languages}}.
\newblock {\slshape \bibinfo{journal}{Foundations and Trends in Programming
  Languages}} \bibinfo{volume}{3}, \doi{10.1561/2500000031}.

\bibitemdeclare{article}{AH2016:JNSM}
\bibitem{AH2016:JNSM}
\bibinfo{author}{Paul \surnamestart Anderson\surnameend} \&
  \bibinfo{author}{Herry \surnamestart Herry\surnameend}
  (\bibinfo{year}{2016}): \emph{\bibinfo{title}{A Formal Semantics for the
  {S}mart{F}rog Configuration Language}}.
\newblock {\slshape \bibinfo{journal}{J. Netw. Syst. Manag.}},
  \doi{10.1007/s10922-015-9351-y}.

\bibitemdeclare{article}{BCR2019:JCSS}
\bibitem{BCR2019:JCSS}
\bibinfo{author}{Yohan \surnamestart Boichut\surnameend},
  \bibinfo{author}{Jacques \surnamestart Chabin\surnameend} \&
  \bibinfo{author}{Pierre \surnamestart R{\'{e}}ty\surnameend}
  (\bibinfo{year}{2019}): \emph{\bibinfo{title}{Towards more precise rewriting
  approximations}}.
\newblock {\slshape \bibinfo{journal}{J. Comput. Syst. Sci.}}
  \bibinfo{volume}{104}, pp. \bibinfo{pages}{131--148},
  \doi{10.1016/J.JCSS.2017.01.006}.

\bibitemdeclare{inproceedings}{chardet:hal-02535077}
\bibitem{chardet:hal-02535077}
\bibinfo{author}{Maverick \surnamestart Chardet\surnameend},
  \bibinfo{author}{H{\'e}l{\`e}ne \surnamestart Coullon\surnameend} \&
  \bibinfo{author}{Christian \surnamestart P{\'e}rez\surnameend}
  (\bibinfo{year}{2020}): \emph{\bibinfo{title}{{Predictable Efficiency for
  Reconfiguration of Service-Oriented Systems with Concerto}}}.
\newblock In: {\slshape \bibinfo{booktitle}{{20th IEEE/ACM International
  Symposium on Cluster, Cloud and Internet Computing (CCGrid)}}},
  \bibinfo{publisher}{{IEEE}}, \doi{10.1109/CCGrid49817.2020.00-59}.

\bibitemdeclare{unpublished}{chardet:hal-02737859}
\bibitem{chardet:hal-02737859}
\bibinfo{author}{Maverick \surnamestart Chardet\surnameend},
  \bibinfo{author}{H{\'e}l{\`e}ne \surnamestart Coullon\surnameend},
  \bibinfo{author}{Christian \surnamestart P{\'e}rez\surnameend},
  \bibinfo{author}{Dimitri \surnamestart Pertin\surnameend},
  \bibinfo{author}{Charl{\`e}ne \surnamestart Servantie\surnameend} \&
  \bibinfo{author}{Simon \surnamestart Robillard\surnameend}
  (\bibinfo{year}{2020}): \emph{\bibinfo{title}{{Enhancing Separation of
  Concerns, Parallelism, and Formalism in Distributed Software Deployment with
  Madeus}}}.
\newblock \bibinfo{note}{{h}al:
  \href{https://inria.hal.science/hal-02737859}{hal-02737859}}.

\bibitemdeclare{article}{CHARDET2021102582}
\bibitem{CHARDET2021102582}
\bibinfo{author}{Maverick \surnamestart Chardet\surnameend},
  \bibinfo{author}{Hélène \surnamestart Coullon\surnameend} \&
  \bibinfo{author}{Simon \surnamestart Robillard\surnameend}
  (\bibinfo{year}{2021}): \emph{\bibinfo{title}{Toward safe and efficient
  reconfiguration with Concerto}}.
\newblock {\slshape \bibinfo{journal}{Sci. Comput. Program.}},
  \doi{10.1016/j.scico.2020.102582}.
\newblock \bibinfo{note}{{h}al:
  \href{https://hal.science/hal-03103714/}{hal-03103714}}.

\bibitemdeclare{article}{CCT2009:FTDB}
\bibitem{CCT2009:FTDB}
\bibinfo{author}{James \surnamestart Cheney\surnameend}, \bibinfo{author}{Laura
  \surnamestart Chiticariu\surnameend} \& \bibinfo{author}{Wang~Chiew
  \surnamestart Tan\surnameend} (\bibinfo{year}{2009}):
  \emph{\bibinfo{title}{Provenance in Databases: Why, How, and Where}}.
\newblock {\slshape \bibinfo{journal}{Found. Trends Databases}},
  \doi{10.1561/1900000006}.

\bibitemdeclare{misc}{clavel2024maude}
\bibitem{clavel2024maude}
\bibinfo{author}{Manuel \surnamestart Clavel\surnameend},
  \bibinfo{author}{Francisco \surnamestart Dur{\'a}n\surnameend},
  \bibinfo{author}{Steven \surnamestart Eker\surnameend},
  \bibinfo{author}{Santiago \surnamestart Escobar\surnameend},
  \bibinfo{author}{Patrick \surnamestart Lincoln\surnameend},
  \bibinfo{author}{Narciso \surnamestart Mart{\i}-Oliet\surnameend},
  \bibinfo{author}{Jos{\'e} \surnamestart Meseguer\surnameend},
  \bibinfo{author}{Rub{\'e}n \surnamestart Rubio\surnameend} \&
  \bibinfo{author}{Carolyn \surnamestart Talcott\surnameend}
  (\bibinfo{year}{2024}): \emph{\bibinfo{title}{Maude manual (version 3.4)}}.
\newblock \urlprefix\url{https://maude.lcc.uma.es/maude-manual/}.

\bibitemdeclare{article}{CHL2023:CSUR}
\bibitem{CHL2023:CSUR}
\bibinfo{author}{H\'{e}l\`{e}ne \surnamestart Coullon\surnameend},
  \bibinfo{author}{Ludovic \surnamestart Henrio\surnameend},
  \bibinfo{author}{Fr\'{e}d\'{e}ric \surnamestart Loulergue\surnameend} \&
  \bibinfo{author}{Simon \surnamestart Robillard\surnameend}
  (\bibinfo{year}{2023}): \emph{\bibinfo{title}{Component-Based Distributed
  Software Reconfiguration: A Verification-Oriented Survey}}.
\newblock {\slshape \bibinfo{journal}{ACM Comput. Surv.}},
  \doi{10.1145/3595376}.

\bibitemdeclare{article}{dicosmo2014ic}
\bibitem{dicosmo2014ic}
\bibinfo{author}{Roberto \surnamestart Di~Cosmo\surnameend},
  \bibinfo{author}{Jacopo \surnamestart Mauro\surnameend},
  \bibinfo{author}{Stefano \surnamestart Zacchiroli\surnameend} \&
  \bibinfo{author}{Gianluigi \surnamestart Zavattaro\surnameend}
  (\bibinfo{year}{2014}): \emph{\bibinfo{title}{{Aeolus}: a Component Model for
  the Cloud}}.
\newblock {\slshape \bibinfo{journal}{{Information and Computation}}},
  \doi{10.1016/j.ic.2014.11.002}.

\bibitemdeclare{misc}{nuitp}
\bibitem{nuitp}
\bibinfo{author}{F~\surnamestart Dur{\'a}n\surnameend},
  \bibinfo{author}{S~\surnamestart Escobar\surnameend},
  \bibinfo{author}{J~\surnamestart Meseguer\surnameend} \&
  \bibinfo{author}{J~\surnamestart Sapina\surnameend} (\bibinfo{year}{2024}):
  \emph{\bibinfo{title}{An Inductive Theorem Prover for Maude Equational
  Theories}}.
\newblock \urlprefix\url{https://nuitp.webs.upv.es/download/NuITP.pdf}.

\bibitemdeclare{inproceedings}{FM2006:WRLA}
\bibitem{FM2006:WRLA}
\bibinfo{author}{Azadeh \surnamestart Farzan\surnameend} \&
  \bibinfo{author}{Jos{\'{e}} \surnamestart Meseguer\surnameend}
  (\bibinfo{year}{2006}): \emph{\bibinfo{title}{Partial Order Reduction for
  Rewriting Semantics of Programming Languages}}.
\newblock In: {\slshape \bibinfo{booktitle}{Workshop on Rewriting Logic and its
  Applications (WRLA)}}, {\slshape \bibinfo{series}{ENTCS}}
  \bibinfo{volume}{176}, \bibinfo{publisher}{Elsevier},
  \doi{10.1016/J.ENTCS.2007.06.008}.

\bibitemdeclare{phdthesis}{FU2019:PHD}
\bibitem{FU2019:PHD}
\bibinfo{author}{Weili \surnamestart Fu\surnameend} (\bibinfo{year}{2019}):
  \emph{\bibinfo{title}{Semantics and provenance of configuration programming
  language $\mu$Puppet}}.
\newblock Ph.D. thesis, \bibinfo{school}{University of Edinburgh, {UK}}.
\newblock
  \urlprefix\url{https://ethos.bl.uk/OrderDetails.do?uin=uk.bl.ethos.798916}.

\bibitemdeclare{inproceedings}{FPA2017:ECOOP}
\bibitem{FPA2017:ECOOP}
\bibinfo{author}{Weili \surnamestart Fu\surnameend}, \bibinfo{author}{Roly
  \surnamestart Perera\surnameend}, \bibinfo{author}{Paul \surnamestart
  Anderson\surnameend} \& \bibinfo{author}{James \surnamestart
  Cheney\surnameend} (\bibinfo{year}{2017}): \emph{\bibinfo{title}{{muPuppet: A
  Declarative Subset of the Puppet Configuration Language}}}.
\newblock In: {\slshape \bibinfo{booktitle}{31st European Conference on
  Object-Oriented Programming (ECOOP)}}, \bibinfo{series}{LIPIcs},
  \doi{10.4230/LIPIcs.ECOOP.2017.12}.

\bibitemdeclare{inproceedings}{GEN1998:RTA}
\bibitem{GEN1998:RTA}
\bibinfo{author}{Thomas \surnamestart Genet\surnameend} (\bibinfo{year}{1998}):
  \emph{\bibinfo{title}{Decidable Approximations of Sets of Descendants and
  Sets of Normal Forms}}.
\newblock In: {\slshape \bibinfo{booktitle}{Rewriting Techniques and
  Applications (RTA)}}, {\slshape \bibinfo{series}{LNCS}}
  \bibinfo{volume}{1379}, \bibinfo{publisher}{Springer}, pp.
  \bibinfo{pages}{151--165}, \doi{10.1007/BFB0052368}.

\bibitemdeclare{inproceedings}{6727828}
\bibitem{6727828}
\bibinfo{author}{Herry \surnamestart Herry\surnameend}, \bibinfo{author}{Paul
  \surnamestart Anderson\surnameend} \& \bibinfo{author}{Michael \surnamestart
  Rovatsos\surnameend} (\bibinfo{year}{2013}):
  \emph{\bibinfo{title}{Choreographing configuration changes}}.
\newblock In: {\slshape \bibinfo{booktitle}{9th International Conference on
  Network and Service Management (CNSM 2013)}},
  \doi{10.1109/CNSM.2013.6727828}.

\bibitemdeclare{article}{KHB2020:JSA}
\bibitem{KHB2020:JSA}
\bibinfo{author}{Khaled \surnamestart Khebbeb\surnameend},
  \bibinfo{author}{Nabil \surnamestart Hameurlain\surnameend} \&
  \bibinfo{author}{Faiza \surnamestart Belala\surnameend}
  (\bibinfo{year}{2020}): \emph{\bibinfo{title}{A {M}aude-Based rewriting
  approach to model and verify {C}loud/{F}og self-adaptation and
  orchestration}}.
\newblock {\slshape \bibinfo{journal}{J. Syst. Archit.}},
  \doi{10.1016/J.SYSARC.2020.101821}.

\bibitemdeclare{article}{KHB2019:IETS}
\bibitem{KHB2019:IETS}
\bibinfo{author}{Khaled \surnamestart Khebbeb\surnameend},
  \bibinfo{author}{Nabil \surnamestart Hameurlain\surnameend},
  \bibinfo{author}{Faiza \surnamestart Belala\surnameend} \&
  \bibinfo{author}{Hamza \surnamestart Sahli\surnameend}
  (\bibinfo{year}{2019}): \emph{\bibinfo{title}{Formal modelling and verifying
  elasticity strategies in cloud systems}}.
\newblock {\slshape \bibinfo{journal}{{IET} Softw.}}
  \bibinfo{volume}{13}(\bibinfo{number}{1}), \doi{10.1049/IET-SEN.2018.5030}.

\bibitemdeclare{inproceedings}{LostanlenBBBFL2021:AM}
\bibitem{LostanlenBBBFL2021:AM}
\bibinfo{author}{Vincent \surnamestart Lostanlen\surnameend},
  \bibinfo{author}{Antoine \surnamestart Bernabeu\surnameend},
  \bibinfo{author}{Jean-Luc \surnamestart B\'{e}chennec\surnameend},
  \bibinfo{author}{Mika\"{e}l \surnamestart Briday\surnameend},
  \bibinfo{author}{S\'{e}bastien \surnamestart Faucou\surnameend} \&
  \bibinfo{author}{Mathieu \surnamestart Lagrange\surnameend}
  (\bibinfo{year}{2021}): \emph{\bibinfo{title}{Energy Efficiency is Not
  Enough:Towards a Batteryless Internet of Sounds}}.
\newblock In: {\slshape \bibinfo{booktitle}{16th International Audio Mostly
  Conference}}, \doi{10.1145/3478384.3478408}.

\bibitemdeclare{book}{olveczky2017distributed}
\bibitem{olveczky2017distributed}
\bibinfo{author}{Peter~Csaba \surnamestart {\"O}lveczky\surnameend}
  (\bibinfo{year}{2017}): \emph{\bibinfo{title}{Designing Reliable Distributed
  Systems}}.
\newblock \bibinfo{publisher}{Springer}, \doi{10.1007/978-1-4471-6687-0}.

\bibitemdeclare{inproceedings}{omond:hal-04372320}
\bibitem{omond:hal-04372320}
\bibinfo{author}{Antoine \surnamestart Omond\surnameend},
  \bibinfo{author}{H{\'e}l{\`e}ne \surnamestart Coullon\surnameend},
  \bibinfo{author}{Issam \surnamestart Ra{\"i}s\surnameend} \&
  \bibinfo{author}{Otto \surnamestart Anshus\surnameend}
  (\bibinfo{year}{2023}): \emph{\bibinfo{title}{{Leveraging Relay Nodes to
  Deploy and Update Services in a CPS with Sleeping Nodes}}}.
\newblock In: {\slshape \bibinfo{booktitle}{{16th IEEE International Conference
  on Cyber, Physical and Social Computing (CPSCom)}}},
  \bibinfo{publisher}{{IEEE}},
  \doi{10.1109/iThings-GreenCom-CPSCom-SmartData-Cybermatics60724.2023.00102}.
\newblock \bibinfo{note}{{h}al:
  \href{https://hal.science/hal-04372320/file/omond_cpscom_2023.pdf}{hal-04372320}}.

\bibitemdeclare{inproceedings}{omond:hal-04372340}
\bibitem{omond:hal-04372340}
\bibinfo{author}{Antoine \surnamestart Omond\surnameend},
  \bibinfo{author}{Issam \surnamestart Ra{\"i}s\surnameend} \&
  \bibinfo{author}{H{\'e}l{\`e}ne \surnamestart Coullon\surnameend}
  (\bibinfo{year}{2023}): \emph{\bibinfo{title}{{Evaluating the energy
  consumption of adaptation tasks for a CPS in the Arctic Tundra}}}.
\newblock In: {\slshape \bibinfo{booktitle}{{19th IEEE International Conference
  on Green Computing and Communications (GreenCom)}}},
  \bibinfo{publisher}{{IEEE}},
  \doi{10.1109/iThings-GreenCom-CPSCom-SmartData-Cybermatics60724.2023.00122}.
\newblock \bibinfo{note}{{h}al:
  \href{https://hal.science/hal-04372340/file/omond_greencom_2023.pdf}{hal-04372340}}.

\bibitemdeclare{inproceedings}{PhilippeOCPR2024:SANER}
\bibitem{PhilippeOCPR2024:SANER}
\bibinfo{author}{Jolan \surnamestart Philippe\surnameend},
  \bibinfo{author}{Antoine \surnamestart Omond\surnameend},
  \bibinfo{author}{Hélène \surnamestart Coullon\surnameend},
  \bibinfo{author}{Charles \surnamestart Prud'Homme\surnameend} \&
  \bibinfo{author}{Issam \surnamestart Raïs\surnameend}
  (\bibinfo{year}{2024}): \emph{\bibinfo{title}{Fast Choreography of
  Cross-DevOps Reconfiguration with Ballet: A Multi-Site OpenStack Case
  Study}}.
\newblock In: {\slshape \bibinfo{booktitle}{IEEE International Conference on
  Software Analysis, Evolution and Reengineering (SANER)}},
  \bibinfo{publisher}{IEEE}.
\newblock \bibinfo{note}{{h}al:
  \href{https://hal.science/hal-04457484}{hal-04457484}}.

\bibitemdeclare{inproceedings}{rubio2022theorem}
\bibitem{rubio2022theorem}
\bibinfo{author}{Rub{\'e}n \surnamestart Rubio\surnameend} \&
  \bibinfo{author}{Adri{\'a}n \surnamestart Riesco\surnameend}
  (\bibinfo{year}{2022}): \emph{\bibinfo{title}{Theorem proving for Maude
  specifications using Lean}}.
\newblock In: {\slshape \bibinfo{booktitle}{International Conference on Formal
  Engineering Methods}}, \bibinfo{organization}{Springer}, pp.
  \bibinfo{pages}{263--280}, \doi{10.1007/978-3-031-17244-1_16}.

\bibitemdeclare{inproceedings}{Sokolowskietal}
\bibitem{Sokolowskietal}
\bibinfo{author}{Daniel \surnamestart Sokolowski\surnameend},
  \bibinfo{author}{Pascal \surnamestart Weisenburger\surnameend} \&
  \bibinfo{author}{Guido \surnamestart Salvaneschi\surnameend}
  (\bibinfo{year}{2021}): \emph{\bibinfo{title}{Automating Serverless
  Deployments for DevOps Organizations}}.
\newblock In: {\slshape \bibinfo{booktitle}{29th ACM Joint Meeting on European
  Software Engineering Conference and Symposium on the Foundations of Software
  Engineering (ESEC/FSE)}}, \doi{10.1145/3468264.3468575}.

\bibitemdeclare{inproceedings}{wildetal}
\bibitem{wildetal}
\bibinfo{author}{Karoline \surnamestart Wild\surnameend}, \bibinfo{author}{Uwe
  \surnamestart Breitenb{\"{u}}cher\surnameend},
  \bibinfo{author}{K{\'{a}}lm{\'{a}}n \surnamestart K{\'{e}}pes\surnameend},
  \bibinfo{author}{Frank \surnamestart Leymann\surnameend} \&
  \bibinfo{author}{Benjamin \surnamestart Weder\surnameend}
  (\bibinfo{year}{2020}): \emph{\bibinfo{title}{Decentralized
  Cross-organizational Application Deployment Automation: An Approach for
  Generating Deployment Choreographies Based on Declarative Deployment
  Models}}.
\newblock In: {\slshape \bibinfo{booktitle}{Advanced Information Systems
  Engineering (CAiSE)}}, \bibinfo{publisher}{Springer},
  \doi{10.1007/978-3-030-49435-3\_2}.

\bibitemdeclare{inproceedings}{YCB2024:FASE}
\bibitem{YCB2024:FASE}
\bibinfo{author}{Geunyeol \surnamestart Yu\surnameend},
  \bibinfo{author}{Seunghyun \surnamestart Chae\surnameend},
  \bibinfo{author}{Kyungmin \surnamestart Bae\surnameend} \&
  \bibinfo{author}{Sungkun \surnamestart Moon\surnameend}
  (\bibinfo{year}{2024}): \emph{\bibinfo{title}{{Formal Specification of
  Trusted Execution Environment APIs}}}.
\newblock In: {\slshape \bibinfo{booktitle}{Fundamental Approaches to Software
  Engineering}}, \bibinfo{publisher}{Springer Nature Switzerland},
  \doi{10.1007/978-3-031-57259-3_5}.

\end{thebibliography}
